\documentclass[aps,prl,floats,twocolumn,showpacs]{revtex4}

\usepackage{graphicx,epsfig}
\usepackage{rotate}
\begin{document}

\title{A mathematical model of turbulent drag reduction by 
high-molecular-weight polymeric additives in a shear flow}

\author{Grigory Isaakovich Barenblatt} 
\affiliation{Department of Mathematics, University of California, Berkeley, CA 94720}


\begin{abstract}

Drag reduction, or, what is the same, mean velocity increase in a turbulent flow at a fixed pressure drop
through the addition of tiny amounts (several parts per million) of high molecular weight polymers (Thoms effect), is known 
already for more than sixty years. Rather long ago it was understood that this effect is related to supramolecular 
structures formed in the flow. Recent experiments by S. Chu, E.S.G. Shaqfeh and their associates, where the motion 
of supramolecular structures was directly observed, made it possible to understand and quantify the dynamic interaction of the 
polymeric structures with the solvent (water) flow. These results lead to the construction of a mathematical model of the Thoms effect,
based on the Kolmogorov(1942)-Prandtl(1945) semi-empirical theory of shear flow turbulence. 
This is the subject of the present Letter.
 \end{abstract}


\maketitle
Let us first introduce the basic equations of the present work.
Turbulent shear flow is considered, so the mean velocity field is assumed to be

\begin{equation}
\overline{u}_1=u(x_2), \hspace{0.3cm} \overline{u}_2=\overline{u}_3=0.
\label{one}
\end{equation}

The density of the solution is indistinguishable from the water density $\rho$, therefore under
zero mean mass force the averaged momentum balance equation has the same form as in the 
absence of the mass force:

\begin{equation}
\overline{T}_{12}= - \rho \overline{u'_{1}u'_{2}}=\tau=\rho u_{\ast}^2.
\label{two}
\end{equation}

In these relations the coordinate $x_1$ is reckoned along the mean flow, the axis $x_2$ is directed perpendicularly to the wall $x_{2}=0$. 
Furthermore, $\overline{T_{12}}$ is the only non-zero component of the Reynolds stress tensor, $\tau$ is the shear stress assumed to be constant across the flow, averaged viscous stress is neglected. The bars denote probability average (mean) values, the turbulent fluctuations are denoted by primes.
The quantity $u_{\ast}=(\tau/\rho)^{1/2}$ is the dynamic, or friction, velocity.

The equation of the turbulent energy balance in the shear flow takes the form:

\begin{equation}
\overline{T}_{12} \frac{du}{dx_2}=  \rho \epsilon - \rho \overline{\bf{u}' \cdot \bf{F}'}.
\label{three}
\end{equation}

Here, unlike in the customary derivations, the mass force is taken into account: $\overline{\bf F}=0$, $\bf{F}'$ is the mass force fluctuation. Furthermore, $\epsilon$ is the mean dissipation rate of the turbulent energy into heat:

\begin{equation}
\epsilon= 2\nu\overline{\bf{D}' \bf{D}'} = \frac{\nu}{2}\overline{(\partial_\alpha u'_\beta +\partial_\beta u'_\alpha)(\partial_\alpha u'_\beta +\partial_\beta u'_\alpha)},
\label{four}
\end{equation}

\noindent $\bf{D}$ is the symmetric part of the strain-rate tensor, $\nu$ is the fluid kinematic viscosity, the summation over repeated Greek indices from 1 to 3 is assumed. The derivation of the equations (\ref{one},\ref{two}) follows the general lines of \cite{monin}.

The turbulent flow is a cascade of vortices of various scales. The basic Kolmogorov's hypothesis \cite{kolmo} (cf. also the Prandtl paper \cite{pra}) can be presented in the following form: at large Reynolds numbers the vortex cascade is self-similar. Hence, all dimensionless flow field properties are universal. According to this hypothesis, all kinematic flow properties are determined by two kinematic properties of different dimensions. Kolmogorov \cite{kolmo} and his direct followers (see \cite{monin}) selected as such quantities the local mean turbulent energy per unit mass $b=(\overline{u'^{2}_{1}+u'^{2}_{2} +u'^{2}_{3}})/2=\overline{u'_{\alpha}u'_{\alpha}}/2$, and the local mean length scale of vortices, or the local external length scale of turbulent flow $l$, which is proportional to it. This hypothesis forms the basis of the $(b,l)$ semi-empirical model. The other possibility is $(b,\epsilon)$ model which also came into wide use. (I used here the original Kolmogorov's notation, often the turbulent energy is denoted by $k$). Furthermore, the momentum exchange coefficient $K$ is introduced:

\begin{equation}
K=\overline{T}_{12}/ (du/dx_2).
\label{five}
\end{equation}

The relation (\ref{five}) is not a new assumption. 

Dimensional analysis suggests the expressions for $K$ and $\epsilon$ (\cite{kolmo}, see also \cite{monin}) via $l$ and $b$:

\begin{equation}
K=l\sqrt{b}, \hspace{0.3cm} \epsilon=\gamma^4 b^{3/2}/l.
\label{six}
\end{equation}

\noindent where $\gamma$ is a universal constant. The estimates in \cite{monin} show, that $\gamma\approx 0.5$. 

Thus we come to the following system of equations for the turbulent shear flow when mean mass force is equal to zero, but the mass force fluctuations are different from zero:

\begin{equation}
l\sqrt{b} \frac{du}{dx_2} = u_{\ast}^{2} 
\label{seven}
\end{equation}

\noindent - the mean momentum balance equation, and

\begin{equation}
l\sqrt{b} (\frac{du}{dx_2})^{2} - \gamma^{4} \frac{b^{3/2}}{l} +  \overline{\bf{u}' \cdot \bf{F}'} =0
\label{eight}
\end{equation}

\noindent - the mean turbulent energy balance equation.

Once the basic equations have been presented we are in place to propose the model of Thoms phenomenon.
Water with polymeric additives which has the property of turbulent drag reduction is not a genuine solution: supramolecular polymeric
structures attaching the solvent molecules are formed in the mixture. The role of such structures in the Thoms phenomenon was suggested 
rather long ago \cite{bare1,bare2}. In the paper \cite{bare2} the photographs of wormlike structures in the carboxymethylcellulose (of rather low molecular weight $\sim 70,000$) water solution were presented. An indirect confirmation of visco-elastic behavior of polymeric supramolecular structures was obtained in \cite{kudin}: a strong water jet directed on a metallic plate did not affect it for hours. Tiny polymer addition leads to a fast abrasive-like wearing of the plate.
However, only in the works by S. Chu, E.S.G. Shaqfeh and their associates the results of direct observations and quantitative experimental measurements of the motions and deformations of the supramolecular structures were obtained (see \cite{sha1,sha2,sha3} and the references presented in these papers). Chu, Shaqfeh and their colleagues observed complicated motions ("tumbling") of supramolecular structures, accompained by their deformation, overturning, rotation, and adjustment of the elongated structures to the local flow. They measured the characteristic time of tumbling, the "disentanglement time" $\theta$.
 
These experimental studies allow to suggest the basic hypothesis of the proposed model:

{\em The force $\bf{F}$ acting on the solvent (water) by the supramolecular polymeric structures, which consist of networks of polymeric molecules with attached solvent molecules, is proportional to the concentrations of the polymers. It is determined also by the instantaneous velocity fluctuation $\bf{u}'=\bf{u}-\overline{\bf{u}}$, and the disentanglement time $\theta$.}

Dimensional analysis and symmetry considerations give

\begin{equation}
{\bf F} = - A s \frac{{\bf u}'}{\theta}
\label{nine}
\end{equation}

the minus sign is due to the obvious fact that the force is a reaction to the velocity fluctuation. Also, A is a dimensionless constant, which could be, in principle, included to the time $\theta$. It was not done here, bearing in mind that the disentanglement time was independently introduced and measured by the experimentalists \cite{sha1,sha2,sha3}.

\begin{figure}[t]
\begin{center}
\epsfig{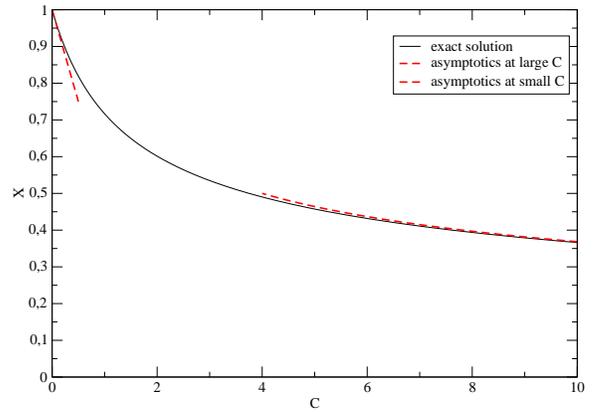}
\end{center} 
  \caption{(color online) Dependence of $X$ on different concentrations $C$ (see text for details).}
\label{f1}
\end{figure}

We summarize: for dilute solutions, when the interaction of supramolecular structures can be neglected, and the density of solutions can be assumed to be equal to the density of the solvent, the first basic equation (\ref{seven}) remains the same as for pure solvent flow.

However, the second basic equation is different. It takes the form:

\begin{equation}
l\sqrt{b} (\frac{du}{dx_2})^{2} - \gamma^{4} \frac{b^{3/2}}{l} - \frac{2 A s b}{\theta} =0
\label{ten}
\end{equation}

\noindent because, obviously, $\overline{\bf{u}'^{2}}=\overline{u'_{\alpha}u'_{\alpha}}=2b$.

This means physically that to the first two terms of the turbulent energy balance equation, representing correspondingly the inflow of the turbulent energy from the mean motion, and dissipation of the turbulent energy into heat, a third term is added, which represents the work spent by turbulence for "disentanglement" of supramolecular structures. Now transform the equation (\ref{ten}) using (\ref{seven}):

\begin{equation}
\frac{u_{\ast}^{4}}{l\sqrt{b}} - \gamma^{4} \frac{b^{3/2}}{l} - \frac{2 A s b}{\theta} =0
\label{eleven}
\end{equation}

Denoting $\gamma b^{1/2}/u_{\ast}$ by $X$, we obtain for $X$ an algebraic equation of the fourth degree:

\begin{equation}
X^4 + 2CX^3 -1 = 0 
\label{twelve}
\end{equation}

\noindent where the dimensionless parameter $C$ is:

\begin{equation}
C=\frac{As}{\gamma^3} \cdot \frac{l}{u_{\ast}\theta}.
\label{thirteen}
\end{equation}

If the polymer concentrations are zero, $C=0$, and we return to the known result: $X=1$; $b=u_{\ast}^{2}/\gamma^{2} = Const$.

For positive $C$, i.e. for a concentration different from zero, $X<1$, $b<u_{\ast}^{2}/\gamma^{2}$. This demonstrates qualitatively, without any assumption concerning $l$, the reduction of the turbulent  energy under the action of polymeric additives. 

We come now to a quantitative model. The length scale $l$ is an independent property, therefore the momentum exchange coefficient $K=l\sqrt{b}$ is also reduced under the action of polymeric additives. According to (\ref{seven}), $du/dx_2 = u_{\ast}^{2}/K$, i.e. the velocity gradient under constant shear stress (constant $u_{\ast}$) is increasing. This leads to an increase of the mean velocity.

Introduce into our consideration a specific expression for the length scale $l$. We accept, as it is done usually, $l=Const \cdot x_2$. According to classic approach $Const=\kappa \gamma$, where $\kappa$ is the von K{\'a}rm{\'a}n constant, $\kappa=0.42$. More refined analysis (see \cite{b1,c2,b3}) gives for $l(x_{2})$ a power law. At large Reynolds numbers, in the intermediate interval of $x_{2}$ the function $l(x_{2})$ can be represented as a linear function of $x_{2}$ so that $Const=\kappa_{\infty}\gamma$, where $\kappa_{\infty}=0.28$. We emphasize that this distinction does not lead for the problem under consideration to any qualitative changes in analysis. So, we assume $l=\kappa_{\infty}\gamma x_2$ so that

\begin{equation}
C=\frac{\kappa_{\infty}As}{\gamma^2} \frac{x_2}{u_{\ast}\theta}.
\label{fourteen}
\end{equation}

The plot of $X=\gamma b^{1/2}/u_{\ast}$ as the function of $C$ is presented in Figure 1; it illustrates the decrease of the turbulent energy. For the momentum exchange coefficient $K=l\sqrt{b}$ we obtain $K=\kappa_{\infty}x_2u_{\ast} X(C)$, and the mean velocity is obtained by integration:

\begin{equation}
u=\frac{u_{\ast}}{\kappa_{\infty}} \int \frac{dx_2}{x_2 X(C)}= \frac{u_{\ast}}{\kappa_{\infty}} \int \frac{dC}{C X(C)} ,
\label{fifteen}
\end{equation}

\noindent because $C$ is, according to (\ref{fourteen}), proportional to $x_2$. It is possible to reduce the last integral to elementary functions. This leads to awkward relations, therefore it is worthwhile to consider the asymptotics, and to compute numerically the result in the intermediate range of C. In the case $C<<1$ (small polymer concentration, small $x_2$, or large disentanglement time $\theta$) equation (\ref{twelve}) gives $X(C)=1-C/2$, so, according to (\ref{fifteen})

\begin{equation}
u=\frac{u_{\ast}}{\kappa_{\infty}} \ln \frac{x_2}{u_{\ast} \theta} + \frac{As}{2\theta\gamma^{2}} x_2 + Const.
\label{sixteen}
\end{equation}

In the case $C>>1$ the first term in (\ref{twelve}) is small in comparison with the second one, and we get $X\sim(2C)^{-1/3}$. This leads to a power law velocity profile:

\begin{equation}
u\sim 3(2\kappa_{\infty}As/\gamma^{2})^{1/3}  u_{\ast}^{2/3} \theta^{-1/3} x_{2}^{1/3} + Const    
\label{seventeen}
\end{equation}

For the intermediate values of the parameter $C$ the second integral in (\ref{fifteen}) can be represented as $\ln C + G(C)+Const$, where

\begin{equation}
G(C)=\int_{0}^{C} \frac{[1-X(C)]}{CX(C)} dC 
\label{eighteen}
\end{equation}

The plot of the function $G(C)$ is presented at the Figure 2. The formulae (\ref{fifteen}), (\ref{eighteen}) allow us to calculate the velocity distribution at arbitrary values of $C$:

\begin{equation}
u=\frac{u_{\ast}}{\kappa_{\infty}} \ln \Big ( \frac{x_2}{u_{\ast} \theta} \Big) + \frac{u_{\ast}}{\kappa_{\infty}} G\Big ( \frac{\kappa_{\infty}As}{\gamma^{2}} \frac{x_2}{u_{\ast} \theta} \Big ) + Const,
\label{nineteen}
\end{equation}

where the constant is determined by the boundary condition at the upper boundary of the viscous sublayer. The $Const$ in equations (\ref{sixteen},\ref{seventeen},\ref{nineteen}) depends on the Reynolds number ($Re$), and at $Re\rightarrow\infty$ it tends to $-\infty$.

\begin{figure}
\begin{center}
\epsfig{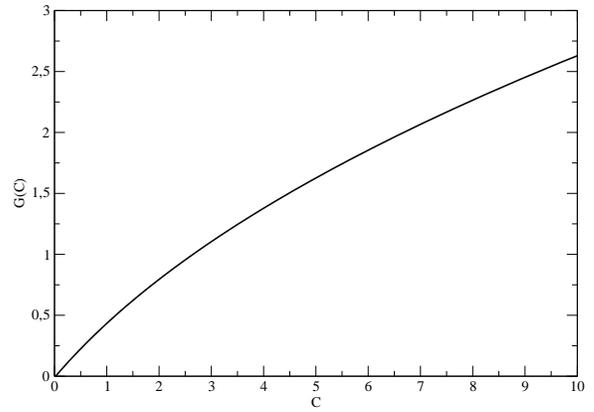}
\end{center} 
  \caption{Integral $G(C)$ as a function of the concentration $C$.}
\label{f2}
\end{figure}

Now we will present a discussion and the conclusions of the current work.
We constructed a semi-emprirical mathematical model of turbulent drag reduction by polymeric additives. It is plausible that the same model will be adequate for shear flows of suspensions of small threads, where the phenomenon of turbulent drag reduction was also observed.

In a discussion of the preliminary version of the present work Professor P. Moin suggested that the decrease of the mean square vorticity $\overline{{\bf w}'^{2}}$, which was also observed in the experimental studies can be explained along the same lines. This statement is correct.
Indeed, the equation for $\bf{\overline{w'^{2}}}$ can be obtained in the same way as the equation for the turbulent energy balance: the operation $curl$ should be applied to the Navier-Stokes equation, the obtained relation for the vorticity multiplied by $2{\bf w}=curl {\bf u}$, and averaged. The next step: the equation for the mean vorticity obtained by the operation $curl$ applied to the Reynolds equation (averaged Navier-Stokes equations) is multiplied by the mean vorticity, and the resulting equation subtracted from the previous one. In this argument instead of the mass force ${\bf F}$ the quantity ${\bf I}=curl {\bf F}$ will enter. Obviously, similarly to the mean value of the mean force ${\bf F}$, the mean value ${\bf I}$ vanishes, due to symmetry. However, there appears an additional term in the equation for $\overline{{\bf w'^{2}}}$ balance: $\overline{{\bf w' I'}}$. Due to the same reasons as in the case of mass force, the quantity ${\bf I}'$ should be proportional to $s{\bf w}'/\theta$:

\begin{equation}
{\bf I}'=-\frac{Bs{\bf w}'}{\theta},
\label{twenty}
\end{equation}

\noindent where $B$ is a constant, analogous to $A$. Therefore in the equation for $\bf{\overline{w'^{2}}}$ an additional term

\begin{equation}
-Bs\overline{\bf{w}'^{2}}/\theta,
\label{twentyone}
\end{equation}

\noindent will appear, analogous to the last term in equation (\ref{eleven}). This demonstrates the reduction of $\bf{\overline{w'^{2}}}$ under the action of the polymeric additives.

The proposed mathematical model of turbulent drag reduction by high molecular weight polymeric additives was constructed under some simplifying assumptions, in particular, constancy of the concentration $s$, and constancy of the disentanglement time $\theta$. Bearing in mind the degree of the accuracy of the Kolmogorov-Prandtl semi-empirical theory of the shear flow turbulence, these assumptions do not seem to be restrictive.

\vspace{0.5cm}
 I want to express my deep gratitude to Professor Marshall Tulin, who drew my attention to the problem of turbulent drag reduction by polymeric additives as far back as 1963 at a remarkable Conference in Tbilissi. In my Moscow group, working on this problem, the motorman was the late experimentalist Dr. V.N. Kalashnikov whom I also remember with deep gratitude. The present work was stimulated by the lecture of Professor E.S.G. Shaqfeh at the Fluid Mechanics Seminar at CTR, Stanford University. I want to express my gratitude to Professor P. Moin for a very fruitful discussion. The attention to the present work of Professor A.J. Chorin is warmly acknowledged. I appreciate the attention and help of Professor A. Arenas.

\end{document}